\documentclass[%
reprint,
superscriptaddress,
nofootinbib,
amsmath,amssymb,
aps,
pra,
showkeys,
]{revtex4-2}

\usepackage[normalem]{ulem} 
\usepackage{url}
\usepackage{xcolor}
\usepackage{hyperref} 
\usepackage[nameinlink,capitalize]{cleveref}
\hypersetup{
	colorlinks = true,
	linkbordercolor = {white},
	linkcolor={purple},
	citecolor={purple},
	urlcolor={blue}
}
\usepackage{comment}
\usepackage{cleveref}

 \usepackage{textcomp}

\usepackage[utf8]{inputenc}
\usepackage{mathtools}
\usepackage{algorithm2e}
\usepackage{graphicx}
\usepackage{dcolumn}
\usepackage{bm}
\usepackage{verbatim}
\usepackage{xcolor}
\usepackage{subcaption}
\usepackage{multirow}

\usepackage{amsmath}

\DeclareMathOperator{\Tr}{Tr}
\definecolor{forestgreen}{rgb}{0.13, 0.55, 0.13}

\usepackage[labelfont=bf,
   justification=Justified,
   format=plain]{caption}

\begin{document}

\title{Learning Time-Varying Gaussian Quantum Lossy Channels}

\author{Angela Rosy Morgillo{\href{https://orcid.org/0009-0006-6142-0692}{\includegraphics[scale=0.004]{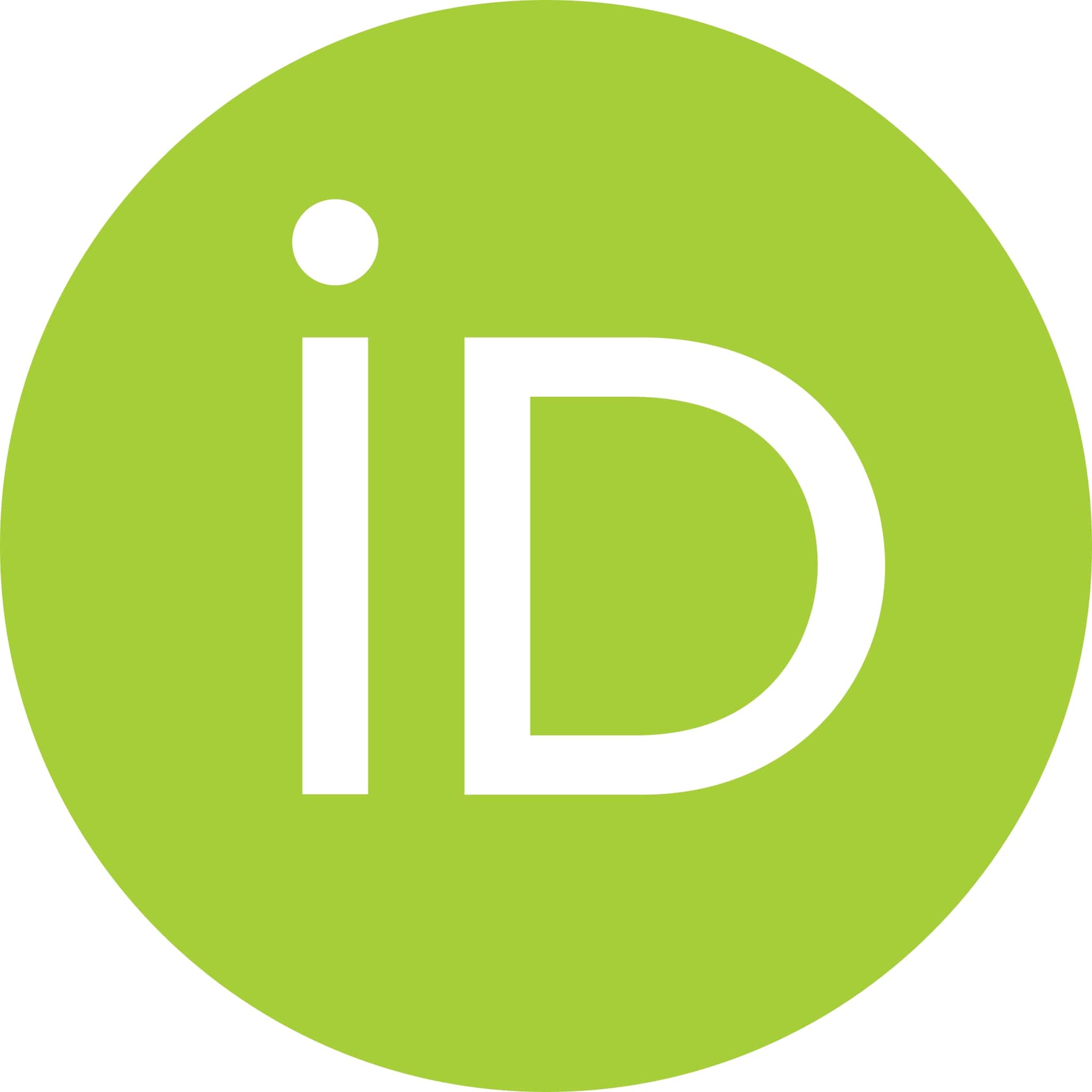}}}}
\email{angelarosy.morgillo01@universitadipavia.it}
\affiliation{Dipartimento di Fisica, Università di Pavia, Via Bassi 6, I-27100, Pavia, Italy}
\affiliation{INFN Sezione di Pavia, Via Bassi 6, I-27100, Pavia, Italy}

\author{Stefano Mancini{\href{https://orcid.org/0000-0002-3797-3987}{\includegraphics[scale=0.004]{images/orcid.jpg}}}}
\affiliation{Scuola di Scienze e Tecnologie, Università di Camerino, I-62032 Camerino, Italy}
\affiliation{INFN Sezione di Perugia, I-06123 Perugia, Italy}

\author{Massimiliano F. Sacchi{\href{https://orcid.org/0000-0002-8909-2196}{\includegraphics[scale=0.004]{images/orcid.jpg}}}}
\affiliation{CNR-Istituto di Fotonica e Nanotecnologie, Piazza Leonardo da Vinci 32, I-20133, Milano, Italy}
\affiliation{Dipartimento di Fisica, Università di Pavia, Via Bassi 6, I-27100, Pavia, Italy}

\author{Chiara Macchiavello{\href{https://orcid.org/0000-0002-2955-8759}{\includegraphics[scale=0.004]{images/orcid.jpg}}}}
\affiliation{Dipartimento di Fisica, Università di Pavia, Via Bassi 6, I-27100, Pavia, Italy}
\affiliation{INFN Sezione di Pavia, Via Bassi 6, I-27100, Pavia, Italy}

\date{\today}

\begin{abstract}
Time-varying quantum channels are essential for modeling realistic quantum systems with evolving noise properties. Here, we consider Gaussian lossy channels varying from one use to another and we employ neural networks to classify, regress, and forecast the behavior of these channels from their Choi-Jamio\l kowski states. The networks achieve at least 87\% of accuracy in distinguishing between non-Markovian, Markovian, memoryless, compound, and deterministic channels. 
In regression tasks, the model accurately reconstructs the loss parameter sequences, and in forecasting, it predicts future values, with improved performance as the memory parameter  approaches 1 for Markovian channels. These results demonstrate the potential of neural networks in characterizing and predicting the dynamics of quantum channels.
\end{abstract}

\keywords{Time-varying channels; Deep Learning; Time series.}

\maketitle

\section{\label{sec:level1}INTRODUCTION}
Quantum communication channels are fundamental to quantum information science, enabling secure communication, quantum cryptography, and distributed quantum computing~\cite{RevModPhys.74.145, caleffi2024distributed}. 

A key challenge is the accurate characterization of quantum channels, which is essential to optimize their use and ensure robust information transfer~\cite{chitambar2022communication}. 

While static quantum channels are well understood, the dynamic nature of time-varying quantum channels makes their characterization significantly more challenging~\cite{3e76bef5-830f-38c1-b8a8-b80a8f60b243, csiszar1988capacity}.

Time-varying quantum channels evolve with each use, with parameters changing either deterministically~\cite{oskouei2021classical}, following a predefined sequence, or randomly, sampled from a probability distribution~\cite{macchiavello2002entanglement, cerf2005quantum, karimipour2006entanglement, demkowicz2007effects, Ruggeri2007, lupo2009forgetfulness, schafer2009capacity}. These channels can be traced back to the framework of quantum memory channels \cite{caruso2014quantum} and are particularly relevant in real-world scenarios where environmental factors or system dynamics shape their properties, playing a crucial role in modeling practical communication systems and analyzing the performance of error correction codes~\cite{etxezarreta2023multiqubit}. 

In this work, we focus on characterizing pure lossy Gaussian quantum channels \cite{holevo2001evaluating} that are time-varying. Lossy Gaussian quantum channels are continuous variable (CV) channels and model the loss of energy en-route. {We choose them not only for their theoretical importance but also because they are well understood and easily implementable in current experimental platforms, making them an ideal testbed for learning-based methods.}

In particular, we focus on estimating the transmissivity parameter $\eta_k \in (0,1)$, which characterizes the lossy nature  of the sequence of channels $\{\mathcal{N}_{\eta_k}\}_{k=1}^\infty$ arising from a time-varying pure lossy Gaussian channel acting on a single bosonic mode. 
 
Our approach involves solving both classification and regression problems. The classification task aims to determine the nature of the lossy parameter \(\eta_k\), which is sampled according to the characteristics of the underlying channel type—non-Markovian, Markovian, memoryless, compound (where the map is fixed but randomly chosen from a set initially), or deterministic.
The regression task involves directly estimating the value of the transmissivity parameter \(\eta_k\). Additionally, we address the forecasting problem, where the goal is to predict future values of the transmissivity parameter based on time series data.

\begin{figure*}[t]
    \centering
    \includegraphics[width=1\textwidth]{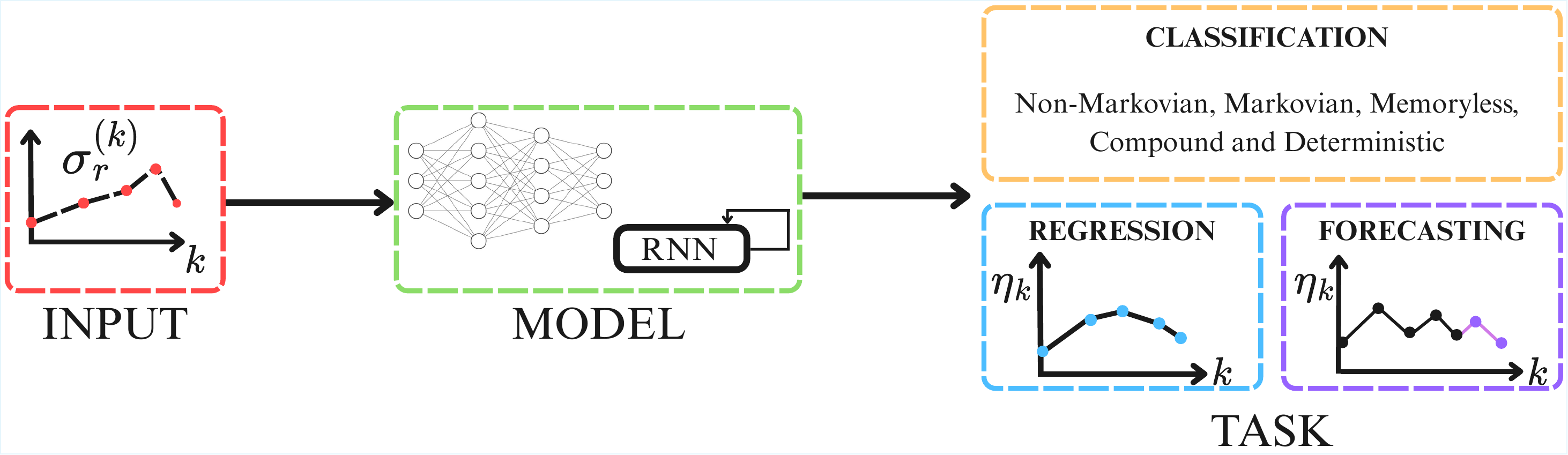}
    \caption{{Schematic representation of the methodology used in this study. The input consists of time series of the first diagonal component of the covariance matrix, depicted in Eq.~(9), associated with consecutive channel uses labeled by \(k\). These inputs are fed into a neural network—such as a feedforward neural network (FFNN), 1D convolutional neural network (1D-CNN), or recurrent neural network (RNN)—to perform three distinct tasks:  
(i) \textbf{Classification}, where the network identifies the type of quantum channel (non-Markovian, Markovian, memoryless, compound, or deterministic) based on the input time series;  
(ii) \textbf{Regression}, aimed at reconstructing the sequence of loss parameters \(\eta_k\) that generated the covariance matrices; and  
(iii) \textbf{Forecasting}, where the network predicts future values of \(\eta_k\) based on previous time steps.
}}
    \label{fig:scheme}
\end{figure*}

{To achieve these objectives, we construct a time series by evaluating, at each channel use, a specific component of the covariance matrix of the Choi-Jamio\l kowski (CJ) state}~\cite{choi1975completely, jamiolkowski1972linear} {associated with the channel, and use this sequence as the input data for neural network-based models.} By analyzing this information using various neural network architectures and methods, we can classify the type of channel, estimate the transmissivity parameter, and forecast its future values. This method thus provides a robust framework for characterizing time-varying quantum channels. The scheme of the methodology used in this work is shown in \cref{fig:scheme}.

{A related approach to model quantum dynamics was presented in} \cite{ostaszewski2019approximation}{, where bidirectional long short-term memory (LSTM) architectures were used to learn corrections to control pulses in a two-qubit system. Although both works leverage recurrent neural networks to analyze quantum time series, our approach differs in several important respects. First, we consider continuous-variable quantum channels rather than finite-dimensional control systems, enabling a more scalable framework applicable to infinite-dimensional Hilbert spaces. Second, we adopt unidirectional recurrent architectures— including LSTMs, one-dimensional convolutional neural networks (1D-CNNs), and feedforward neural networks— selected based on performance and computational efficiency. Finally, the goals of the two studies are distinct: while} \cite{ostaszewski2019approximation} {focuses on quantum control, our work addresses the classification of channel types and the prediction of time-varying channel parameters.}

This paper is structured as follows. \cref{sec:theory} provides the theoretical background, including an overview of Gaussian channels and relevant mathematical tools. \cref{Sec:models} introduces the specific channels considered in this work and the modeling of the lossy parameter. \cref{sec:results} presents the results, including the application of neural networks and the construction of the dataset, which are explained in the context of each specific task. Finally, \cref{sec:conclusion} discusses the implications of our findings and outlines future research directions.

\section{Gaussian Lossy Channels}
\label{sec:theory}
In this section, we introduce Gaussian lossy channels and outline their key mathematical properties~\cite{serafini2023quantum}. Particular attention is paid to the form of the covariance matrix associated with the CJ states of these channels, as it serves as the foundation for building the dataset used in this work.

Consider the vector of canonical operators for a single-mode bosonic system, defined as \(\hat{\boldsymbol{r}} = (\hat{x}, \hat{p})^T\). A Gaussian state \(\rho_G\) is fully characterized by its first two statistical moments: the displacement vector \(\bar{\boldsymbol{r}}\) and the covariance matrix \(\boldsymbol{\sigma}\). These are given by:  
\begin{align}  
    \bar{\boldsymbol{r}} &= \Tr[\hat{\boldsymbol{r}} \rho_G], \\  
    \boldsymbol{\sigma} &= \Tr\left[
    \{ (\hat{\boldsymbol{r}} - \bar{\boldsymbol{r}}), (\hat{\boldsymbol{r}} - \bar{\boldsymbol{r}})^T  \}\rho_G\right], 
    \label{eq:moments}  
\end{align} 
where $\{\hat{b}, \hat{b}^T\} = \hat{b} \hat{b}^T + (\hat{b}\hat{b}^T)^T$. \(\bar{\boldsymbol{r}}\) represents the expectation values of the quadrature operators, and \(\boldsymbol{\sigma}\) captures the quantum fluctuations and correlations between them. 

Next, we describe how a Gaussian lossy channel acts on the mean value \( \bar{\boldsymbol{r}} \) and covariance matrix \( \boldsymbol{\sigma} \) of a quantum state. Specifically, a Gaussian lossy channel introduces loss in the system, characterized by the lossy parameter \( \eta \in [0, 1] \), and may also introduce thermal noise. The transformations governing the evolution of the mean and covariance are as follows:
\begin{align}
    \bar{\boldsymbol{r}} &\longrightarrow X\bar{\boldsymbol{r}}, 
    \notag\\
    \boldsymbol{\sigma} &\longrightarrow X \boldsymbol{\sigma} X^T + Y,
    \label{eq:meanvar}
\end{align}
where \( X \) is a matrix that depends on the parameter \( \eta \), and \( Y \) represents the added thermal noise.

For simplicity, we assume a pure lossy channel where the matrices \( X \) and \( Y \) are defined as follows:
\begin{align}
X &= \sqrt{\eta} I_2, \\
Y &= (1-\eta) I_2,
\end{align}
with \( I_2 \) denoting the \( 2 \times 2 \) identity matrix. This assumption allows us to focus on the primary effects of loss, namely considering a quantum-limited attenuator with minimal added noise.

This approach can be extended to the Choi–Jamio\l{}kowski representation~\cite{choi1975completely, jamiolkowski1972linear}, which provides a bijective correspondence between completely positive (CP) maps on the space of bounded operators \( \mathcal{B}(L^2(\mathbb{R})) \) and density operators on \( L^2(\mathbb{R}^2) \). Within this representation, the effect of a Gaussian lossy channel can be analyzed through its impact on the covariance matrix of a two-mode entangled state. 

Consider the two-mode Gaussian entangled state
\begin{equation}
|\psi_r \rangle = \frac{1}{\cosh{r}}\sum_{j=0}^\infty (\tanh{r})^j |j \rangle |j \rangle,
\label{eq:twomodestate}
\end{equation}
where \( r \) controls the degree of entanglement, with \( r \to \infty \) corresponding to maximal entanglement. The associated covariance matrix, given by  
\begin{equation}
\boldsymbol{\sigma}_r = \begin{pmatrix}
    \cosh{2r} I_2 & \sinh{2r} \Sigma \\
    \sinh{2r} \Sigma & \cosh{2r} I_2
\end{pmatrix}, \quad \Sigma = \begin{pmatrix}
    1 & 0 \\ 
    0 & -1
\end{pmatrix},
\label{eq: generalmoments}
\end{equation}
encodes the quantum correlations between the modes, while the displacement vector \( \bar{\boldsymbol{r}} = 0 \) reflects the absence of coherent displacements in phase space.  

The CJ state for a given \( r \) is defined as  
\begin{equation}
\hat{\phi}_r = (\mathcal{N} \otimes id)|\psi_r\rangle \langle \psi_r|,
\label{eq:choi}
\end{equation}  
where \( \mathcal{N} \) is the CP map acting on one mode, and \( id \) is the identity map on the other mode. Notably, the infinite Schmidt rank of \( |\psi_r\rangle \) ensures this isomorphism remains valid even for finite \( r \), allowing for a faithful representation of the map \( \mathcal{N} \)~\cite{d2005characterization}. While the choice of a maximally entangled state is conventional, it is not strictly required to maintain the bijective correspondence between maps and states.  

As shown in \cref{eq:meanvar}, the action of a Gaussian lossy channel on a Gaussian state leads to an output that is fully characterized by its covariance matrix. In the case of a time-varying quantum channel, where the loss parameter \(\eta_k\) changes across discrete uses, the covariance matrix of the output state at the \(k\)-th use is given by:
\begin{widetext}
\begin{equation}
\begin{split}
 \boldsymbol{\sigma}_r^{(k)} = \begin{pmatrix}
        \eta_k \cosh({2r}) + (1 -\eta_k) & 0 &\sqrt{\eta_k} \sinh{2r} & 0 \\
        0 & \eta_k \cosh{2r} + (1 - \eta_k) & 0 & -\sqrt{\eta_k} \sinh{2r} \\
        \sqrt{\eta_k}\sinh{2r} & 0 & \cosh{2r} & 0 \\
        0 & -\sqrt{\eta_k}\sinh{2r} & 0 & \cosh{2r}
    \end{pmatrix}.
\end{split}
\label{eq:covmatrixk}
\end{equation}
\end{widetext}

\section{\label{Sec:models} Channels Types and Sampling}
{In this section, we describe how the time-varying loss parameter \(\eta_k\) is sampled for our simulations. To model the loss parameter, we employed a Beta distribution}~\cite{johnson1994beta}, chosen for its flexibility and its natural support on the interval \([0,1]\), which matches the physical range of \(\eta_k\). The parameters \(\alpha\) and \(\beta\) govern the shape of the distribution, allowing us to represent a wide range of loss behaviors. Importantly, this choice inherently ensures that the sampled values lie within the correct bounds, eliminating the need for any normalization or additional preprocessing of the data. Specifically, the Beta distribution is defined as 
\begin{equation}
    p(x) = \frac{x^{\alpha - 1}(1-x)^{\beta - 1}}{B(\alpha, \beta)},
    \label{eq:betad}
\end{equation}
where $0 \leq x \leq 1$, and $\alpha, \beta > 0$ are the shape parameters, with $B(\alpha, \beta)$ being a normalization factor.\footnote{In our approach, we constrain $\alpha$ and $\beta$ to be greater than 1 to guarantee that the Beta distribution remains uni-modal.}

The mean and variance of this distribution are given by:
\begin{equation}
    \bar{x} = \frac{1}{1 + \frac{\beta}{\alpha}}, \,\,\,\,\,\, \sigma^2 = \frac{\alpha \beta}{(\alpha + \beta)^2(\alpha + \beta + 1)}.
    \label{eq:meanvarbeta}
\end{equation}
Assuming that $\sigma^2 < \bar{x}(1-\bar{x})$, these expressions allow for the computation of the shape parameters \(\alpha\) and \(\beta\) in terms of the mean \(\bar{x}\) and variance \(\sigma^2\), given by:
\begin{equation}
    \alpha = \bar{x} \Big(\frac{\bar{x}(1-\bar{x})}{\sigma^2}-1 \Big), 
    \label{eq:alpha}
\end{equation}
and 
\begin{equation}
  \beta = (1-\bar{x}) \Big(\frac{\bar{x}(1-\bar{x})}{\sigma^2}-1\Big).
    \label{eq:beta}
\end{equation}
This general formulation allows us to define different types of quantum channels by specifying how $\eta_k$ evolves over time. 

After sampling \(\eta_1\) from the Beta distribution  
\begin{equation}
    p(\eta_1) = \frac{\eta_1^{\alpha_1 - 1} (1 - \eta_1)^{\beta_1 - 1}}{B(\alpha_1, \beta_1)},
    \label{eq:priorprob}
\end{equation}  
where \(\alpha_1\) and \(\beta_1\) are the shape parameters corresponding to a given mean \(\bar{\eta}_1\) and variance \(\sigma_1^2\), the subsequent values \(\eta_k\) (\(k \geq 2\)) are drawn from a conditional Beta distribution:  
\begin{equation}
    P(\eta_k | \{\eta_{k-i}\}^{k-1}_{i=1}) = \frac{\eta_k^{\alpha_k - 1} (1 - \eta_k)^{\beta_k - 1}}{B(\alpha_k, \beta_k)},
    \label{eq:condprob}
\end{equation}  
where $\alpha_k$ and $\beta_k$ are derived through Eqs.\eqref{eq:alpha} and \eqref{eq:beta} from the mean and variance evolving at each step as:  
\begin{align}
    \bar{\eta}_k &= \sum_{i=1}^{k-1} \mu_i \eta_{k-i} + \left( 1 - \sum_{i=1}^{k-1} \mu_i \right) \bar{\eta}_1, \notag \\
    \sigma^2_k &= \left( 1 - \sum_{i=1}^{k-1} \mu_i \right) \sigma_1^2, 
    \label{eq:updatememory}
\end{align}
Here, \(\vec{\mu} = \{ \mu_i\}_{i=1}^{k-1}\) is a weight vector that determines how the $k-1$ past values influence the current state, with different choices leading to different memory structures. The vector \(\vec{\eta}_{k-1} = \{\eta_{k-i}\}_{i=1}^{k-1}\) represents the past \(k-1\) values of \(\eta_k\), with the most recent values typically having higher weights.

Based on this framework, we define five models for the evolution of the transmissivity parameter: non-Markovian, Markovian, memoryless, compound, and deterministic channels.  

\paragraph*{Non-Markovian Channel.---} In order to model non-Markovian memory effects, we allow the current state to depend on the previous three values of the transmissivity parameter. Specifically, we assign \(\vec{\mu} = (\mu,\, \mu/2,\, \mu/3)\), giving greater weights to more recent values. To ensure a meaningful level of memory, we set \(\mu \geq 0.2\) in our simulations.

\paragraph*{Markovian Channel.---} In a Markovian channel, each value \(\eta_k\) (\(k \geq 2\)) depends only on the previous value \(\eta_{k-1}\), ensuring that the process follows the Markov property. To maintain a minimal degree of memory, we set \(\vec{\mu} = \mu > 0.1\) in our simulations.

\paragraph*{Memoryless channel.---} For a memoryless channel, the current state of the system is independent of its previous states. Consequently, we set the memory parameter \(\vec{\mu} = 0\) in \cref{eq:updatememory}, ensuring that the mean \(\bar{\eta}_k\) and the variance \(\sigma^2_k\) (as well as the shape parameters \(\alpha_k\) and \(\beta_k\)) remain constant throughout the sequence. As a result, each \(\eta_k\) is drawn independently from the same probability distribution, given in \cref{eq:priorprob}. This ensures that each \(\eta_k\) is sampled from an identical Beta distribution, and no temporal correlations exist between successive samples.

\paragraph*{Compound channel.---} In a compound channel the transmission properties remain fixed over time, meaning that once the initial value \(\eta_1\) is determined, it is kept constant for all subsequent uses of the channel. This implies that \(\eta_k = \eta_1\) for \(k \geq 2\), and no variation occurs in the parameter across the sequence. To model this situation, we use \cref{eq:updatememory} with \(\vec{\mu} = \mu = 1\), which ensures the constancy of \(\eta_k\) over time. In this case, the conditional probability for each element in the sequence is given by:
\begin{equation}
    p(\eta_k | \eta_{k-1}) = \delta( \eta_k - \eta_{k-1}),
    \label{eq:constant}
\end{equation}
where \(\delta\) is the Dirac delta function.

\paragraph*{Deterministic channel.---} A deterministic channel is one where the transmission parameter \(\eta_k\) evolves in a specific and deterministic way over time. In this case, we set both \(\vec{\mu} = 0\) and \(\sigma_1 = \sigma_k = 0\), ensuring that the parameter \(\eta_k\) takes fixed values at each time step. The prior and conditional probabilities for \(\eta_1\) and \(\eta_k\) are given by Dirac delta functions:
\begin{align}
    p(\eta_1 ) &= \delta(\eta_1 - \bar{\eta}_1), \\
    p(\eta_k | \eta_{k-1}) &= \delta (\eta_k - \bar{\eta}_k).
    \label{eq:deterpriori}
\end{align}
We consider two specific forms for \(\bar{\eta}_k\), each describing a different type of variation:
\begin{equation}
    \bar{\eta}_k = a + b e^{\frac{-(k-1)^2}{\Delta}}, 
\end{equation}
and
\begin{equation}
    \bar{\eta}_k = a + b\left|\cos{\frac{k-1}{\Delta}}\right|,
\end{equation}
where \(0 < a,b < \frac{1}{2}\). For the cosine-like modulation, we take \(1 < \Delta < 10\), while for the exponential decay, \(10 < \Delta < 30\) to avoid overly rapid convergence. These models have already been put forward in \cite{oskouei2021classical} as contrasting examples where classical capacity can or cannot be drawn from the limiting behavior of the lossy parameter.
\paragraph*{Dataset Generation.---} The dataset samples are generated following two alternative procedures, \(D_1\) and \(D_2\), which differ in how the initial Beta distribution parameters are chosen. Subsequently, the transmissivity values \(\eta_k\) are sampled sequentially by updating the distribution parameters according to the underlying quantum channel model. The steps are as follows:
\begin{enumerate}
  \item \textbf{Initial Parameter Sampling (D\textsubscript{1} or D\textsubscript{2}):}
  In the first approach (D\textsubscript{1}), shape parameters \(\alpha_1, \beta_1\) are randomly sampled from the given interval $[1, 10]$. \\
  Alternatively, in the second approach (D\textsubscript{2}), we directly sample the mean \(\bar{\eta}_1\) and variance \(\sigma_1^2\) of the first Beta distribution.
  \item \textbf{Sampling \(\eta_1\):}
  Using the parameters \(\alpha_1, \beta_1\) (from D\textsubscript{1}) or \(\bar{\eta}_1, \sigma_1^2\) (from D\textsubscript{2}), define the Beta distribution \(B(\alpha_1, \beta_1)\) and sample the first transmissivity value \(\eta_1\).
  \item \textbf{Recursive Updates for \(k \geq 2\):} For each subsequent time step \(k \geq 2\), we update the mean \(\bar{\eta}_k\) and variance \(\sigma_k^2\) according to the update rules determined by the chosen quantum channel type. \\
  We then construct the Beta distribution \(B(\alpha_k, \beta_k)\) corresponding to the updated mean and variance. We finally sample \(\eta_k\) from this Beta distribution.
\end{enumerate}

\section{Results}
\label{sec:results}
We present results from simulations designed to address three distinct problems using machine and deep learning models. Simulations were performed with TensorFlow~\cite{tensorflow2015-whitepaper}, and training was performed using Adam optimizer~\cite{kingma2014adam}. The dataset includes discrete time series of covariance matrices, as defined in \cref{eq:covmatrixk}, representing multiple uses of a quantum channel. The main objectives of this study are: (i) to evaluate the network ability to classify different types of quantum channels based on the time series of covariance matrices; (ii) to assess the network performance in reconstructing the lossy parameters, \(\eta_k\), that generated the observed sequences; and (iii) to investigate the network ability to predict future values of the lossy parameters based on the input time series. 

\subsection{Classification} Here the goal is to classify quantum channels into five distinct categories based on time series data extracted from their covariance matrices. Specifically, the channels are categorized as non-Markovian (NM), Markovian (M), memoryless (ML), compound (C), or deterministic (D). The input of the dataset $\{\boldsymbol{\sigma}^{(k)}_r\}_{k=1}^{N}$ consists of a discrete time series comprising $N$ elements, where each element represents the first diagonal component of the covariance matrix (\ref{eq:covmatrixk}) associated with the $k$-th use of the channel. We chose to focus solely on this component because it contains sufficient information for classification and helps accelerate the training process. The output is a categorical class label $\boldsymbol{y}$ indicating the type of channel. 

To generate the dataset, we consider two different approaches. In the first approach ($D_1$), we directly sample the first element of each time series from a Beta distribution with shape parameters $1 \leq \alpha, \beta \leq 10$, chosen uniformly. In the second approach ($D_2$), we first sample a mean $0 \leq \bar{\eta}_1 \leq 1$ and a variance $0 \leq \sigma_1^2 \leq \min\left( \frac{\bar{\eta}_1(1 - \bar{\eta}_1)}{1 + 1/\bar{\eta}_1}, \, \frac{\bar{\eta}_1(1 - \bar{\eta}_1)}{1 + 1/(1 - \bar{\eta}_1)} \right)$, ensuring that the resulting Beta distribution has $\alpha, \beta > 1$. The second option is less biased, as it allows for a broader range of possible distributions and introduces a more complex problem to solve; in particular, it enables us to sample the transmissivity parameter $\eta_1$ uniformly across its entire possible range.
\begin{figure*}[t]
    \centering
    \subfloat[\label{subfig:a1}]{{\includegraphics[trim={0.6cm 0.6cm 0.2cm 0.6cm},clip, width=0.45 \textwidth]{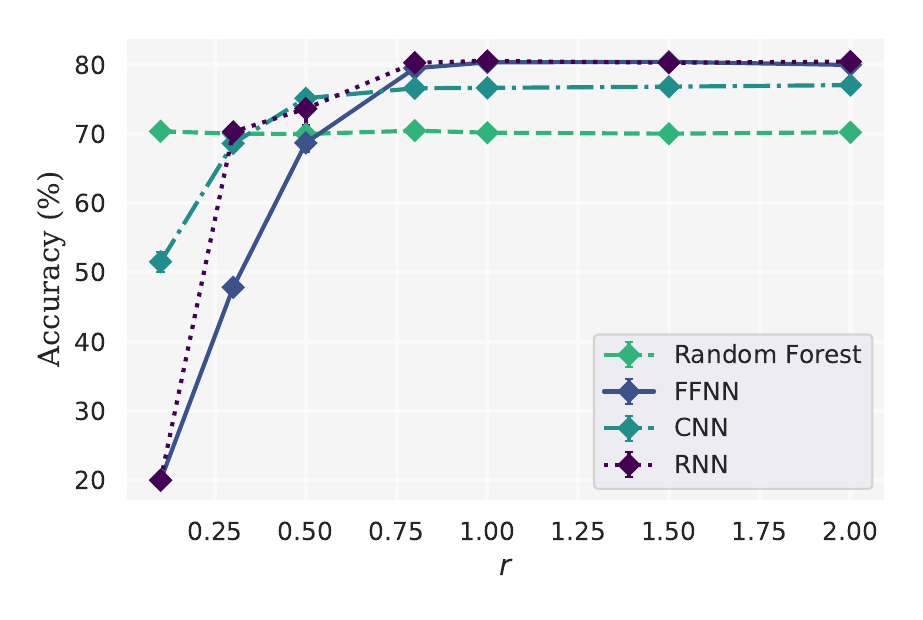} }}%
    \subfloat[\label{subfig:a2}]{{\includegraphics[trim={0.6cm 0.6cm 0.2cm 0.6cm},clip, width=0.45 \textwidth]{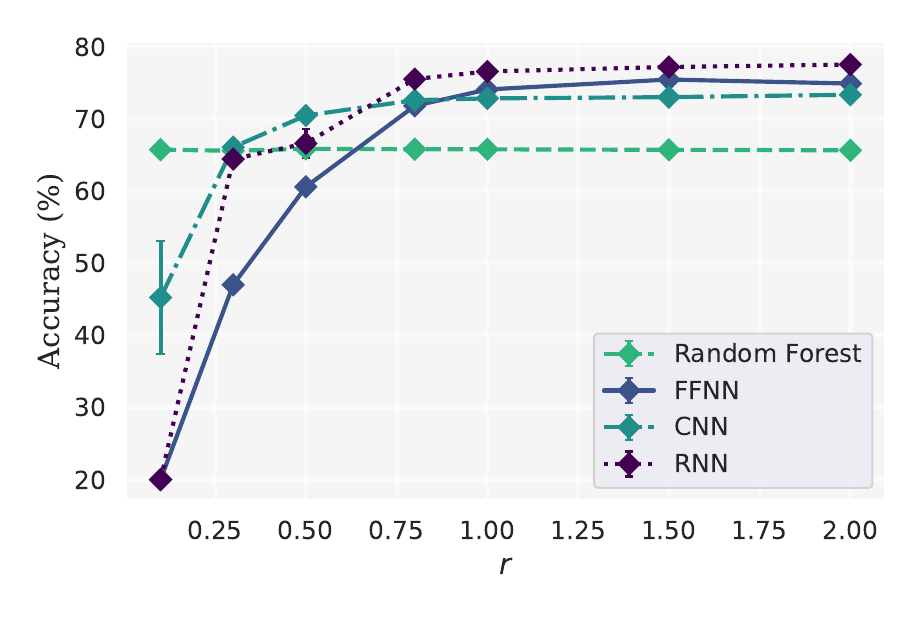} }}\\
    \subfloat[\label{subfig:a3}]{{\includegraphics[trim={0.6cm 0.6cm 0.2cm 0.6cm},clip, width=0.45 \textwidth]{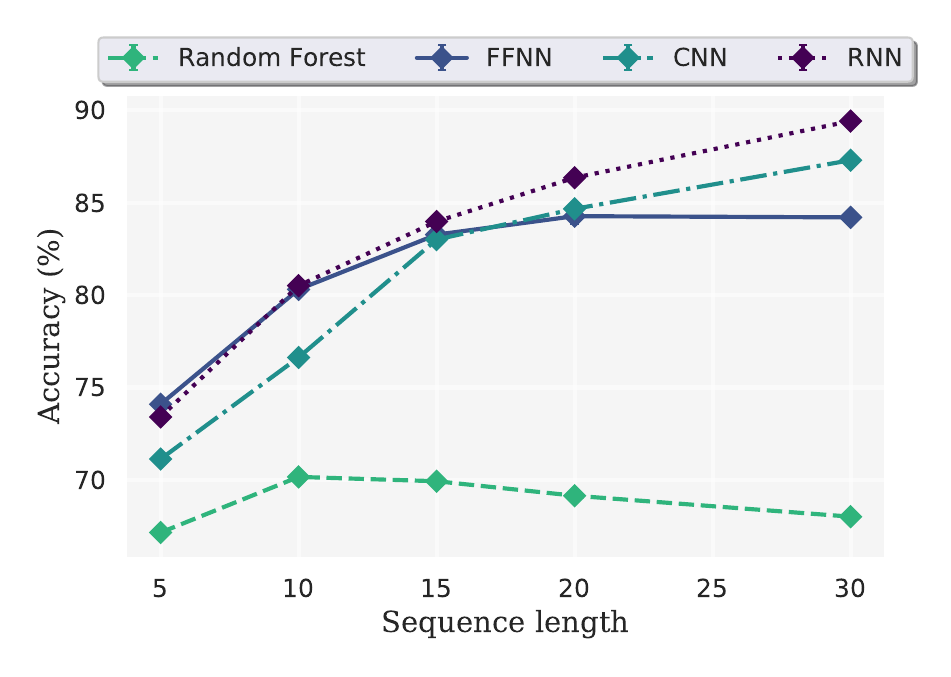} }}%
    \subfloat[\label{subfig:a4}]{{\includegraphics[trim={0.6cm 0.6cm 0.2cm 0.6cm},clip, width=0.45 \textwidth]{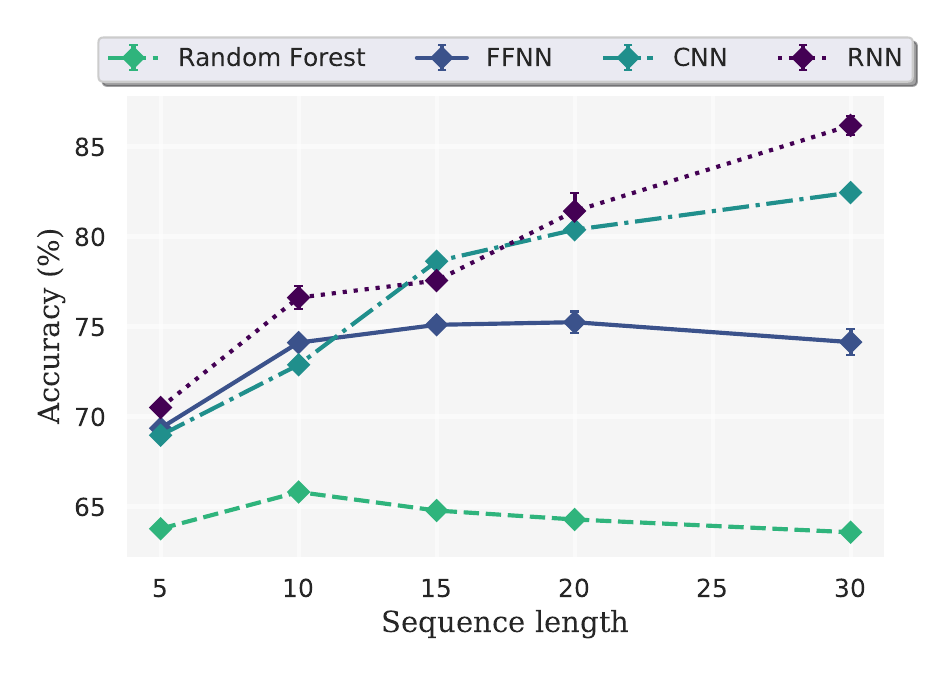} }}%
    \caption{Classification accuracy of the different models analyzed: Random Forest, Feedforward Neural Network (FFNN), Recurrent Neural Network (RNN), and 1D-Convolutional Neural Network (1D-CNN). (a) Accuracy as a function of the entanglement parameter $r$ for the dataset generation approach $D_1$. (b) Same as (a), but for $D_2$. (c) Accuracy as a function of the sequence length for $D_1$. (d) Same as (c), but for $D_2$. Each data point is obtained by averaging the accuracy over 5 independent runs, selecting in each run the model with the highest test accuracy. Error bars represent the standard deviation across these runs.}
    \label{fig:acc}
\end{figure*}
For this classification task, we explore four different techniques:  
\begin{itemize}
    \item \textbf{Random Forest}: A widely used ensemble learning method that constructs multiple decision trees and aggregates their outputs~\cite{biau2016random}.  
    \item \textbf{Feedforward Neural Networks (FFNNs)}: Standard deep learning models where information flows in one direction, from input to output~\cite{svozil1997introduction}.  
    \item \textbf{Recurrent Neural Networks (RNNs)}: Neural networks designed for sequential data, capable of capturing temporal dependencies by maintaining a hidden state that evolves over time. In particular, we employed Long Short-Term Memory (LSTM) networks~\cite{salehinejad2017recent}, which mitigate vanishing gradient issues and are well-suited for learning long-range dependencies.    
    \item \textbf{1D-Convolutional Neural Networks (1D-CNNs)}: Convolutional networks applied to one-dimensional sequences, particularly effective for time series classification as they can extract local patterns and temporal correlations from the data~\cite{kiranyaz20211d}.  
\end{itemize}  
Details regarding the architectures and hyperparameters used can be found in \cref{app:architectures}.

For all models, we used a dataset of 50000 samples (10000 per class), split into training and test sets with an 80:20 ratio. The performance of these models is analyzed in \cref{fig:acc}, where we report the classification accuracy –defined as the fraction of correctly predicted samples over the total number of predictions—against the entanglement parameter $r$ and the sequence length, for both dataset generation strategies.  

Each accuracy value is averaged over five runs, selecting in each run the model with the highest test accuracy. This allows us to assess the impact of randomness in both dataset generation and parameters initialization. We observe that the performance of RNNs and 1D-CNNs improves as the squeezing parameter \(r\) increases, whereas Random Forest accuracy remains flat. Increasing the sequence length benefits both RNNs and 1D-CNNs, which are designed to learn from sequential dependencies. FFNNs also exhibit performance improvements with longer sequences, though to a lesser extent, as they are not explicitly designed to capture temporal correlations like RNNs and 1D-CNNs. In contrast, the Random Forest classifier---which lacks temporal modeling capabilities---performs poorly in this setting, as the model treats each sequence as a static feature vector, ignoring the temporal order of the inputs. Therefore, increasing the sequence length only extends this single feature without adding new independent information, leading to flat performance.
While there exist variants of Random Forest, such as those using block bootstrapping~\cite{goehry2023random}, which can better handle temporal dependencies, these methods rely on continuous and autocorrelated time series. Our dataset consists of many independent sequences without temporal continuity between them, so such variants are not suitable here. In contrast, recurrent models such as RNNs are explicitly designed to capture temporal dependencies, and thus their performance scales more effectively with sequence length. We ensured all models had roughly the same number of trainable parameters for a fair comparison. Even so, RNNs consistently performed best, showing they are better suited for capturing temporal dependencies.
Finally, we confirm that the second dataset generation approach $D_2$ introduces a more challenging classification problem, resulting in slightly lower accuracy and increased performance fluctuations. 

\begin{figure}[]
    \centering
    \begin{subfigure}[b]{8.5cm}
        \centering
        \includegraphics[trim={0cm 0.2cm 0 0},clip,width=0.9\linewidth]{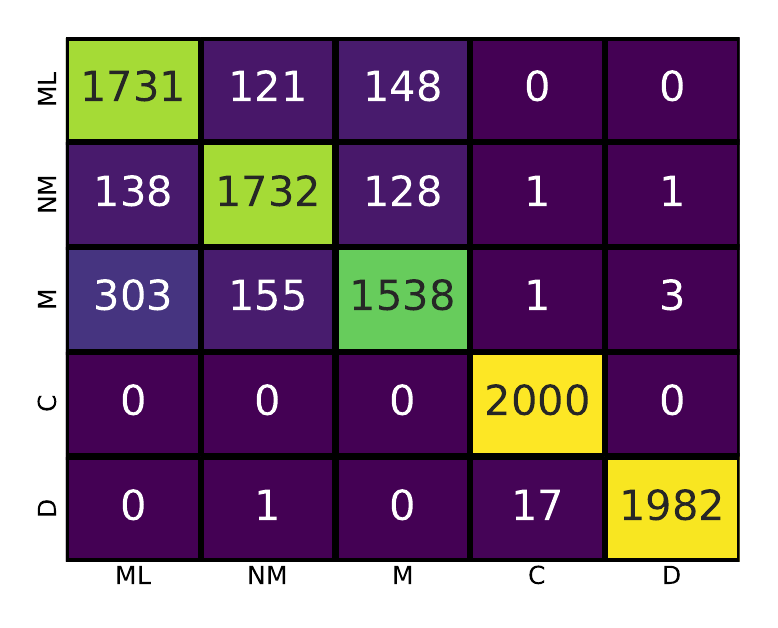}
        \caption{Best classification performance on the $D_1$ dataset.}
        \label{fig:subfig1}
    \end{subfigure}
    \vspace{-0.2\baselineskip} 
    \begin{subfigure}[b]{8.5cm}
        \centering
        \includegraphics[trim={0cm 0.2cm 0 0},clip,width=0.9\linewidth]{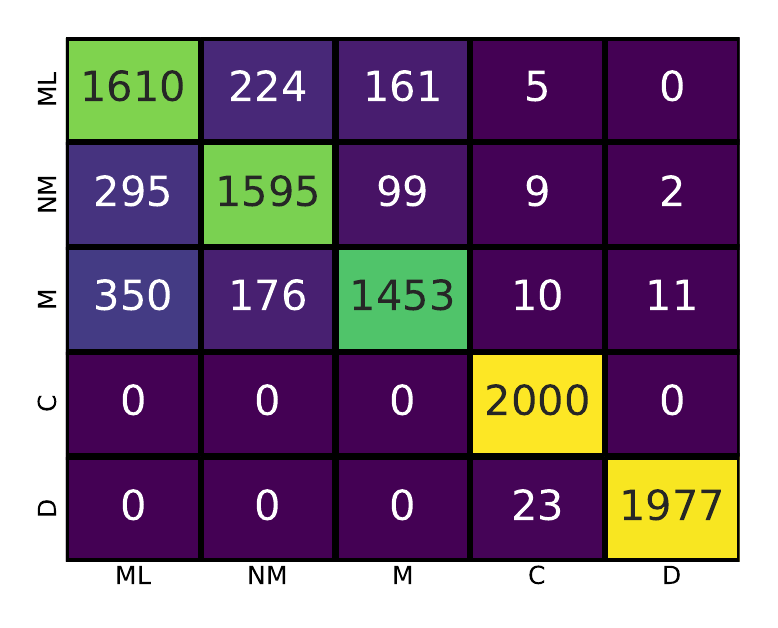}
        \caption{Best classification performance on the $D_2$ dataset.}
        \label{fig:subfig2}
    \end{subfigure}
    \caption{Confusion matrices for the best classification performances of both approaches. (a) For $D_1$, and (b) for $D_2$. The rows represent the true class labels, while the columns show the predicted labels. The best results are achieved using RNNs with a sequence length of 30, with accuracy rates of 89.83\% for $D_1$ and 87.11\% for $D_2$.}
    \label{fig:confmats}
\end{figure}

The best classification performances for both approaches, $D_1$ and $D_2$, are presented in \cref{fig:confmats}, which displays the corresponding confusion matrices. These matrices illustrate the distribution of true versus predicted labels, providing insight into the model classification accuracy and error patterns. For both approaches, the optimal results are obtained using RNNs with a sequence length of 30 and entanglement parameter $r=1$.\footnote{Comparable results can be achieved with $r>1$, as shown in \cref{fig:acc}.}
For $D_1$, the highest accuracy achieved is approximately 90\%, while for $D_2$, it is 87\%. 
While the model demonstrates near-perfect classification for classes C (compound) and D (deterministic), it exhibits challenges in distinguishing between classes M (Markovian) and ML (memoryless). The first diagonal components of the covariance matrix, which serve as the input features, may exhibit similar statistical patterns for Markovian and memoryless channels, especially for small values of the memory parameter $\mu$, making it harder for the model to differentiate them.
\paragraph*{Memory Analysis.---} To better understand why Markovian channels are frequently misclassified as memoryless, we analyze how the memory parameter \(\mu\) influences the degree of Markovianity. Specifically, we frame a binary classification task, where each class corresponds to a range of \(\mu\) values. We defined the bin edges as \([0, c, 1]\), with $c$ varying between 0.3 and 0.96.
\begin{figure}
    \centering 
    \includegraphics[trim={0.6cm 0.2cm 0.4 0},clip,width=1\linewidth]{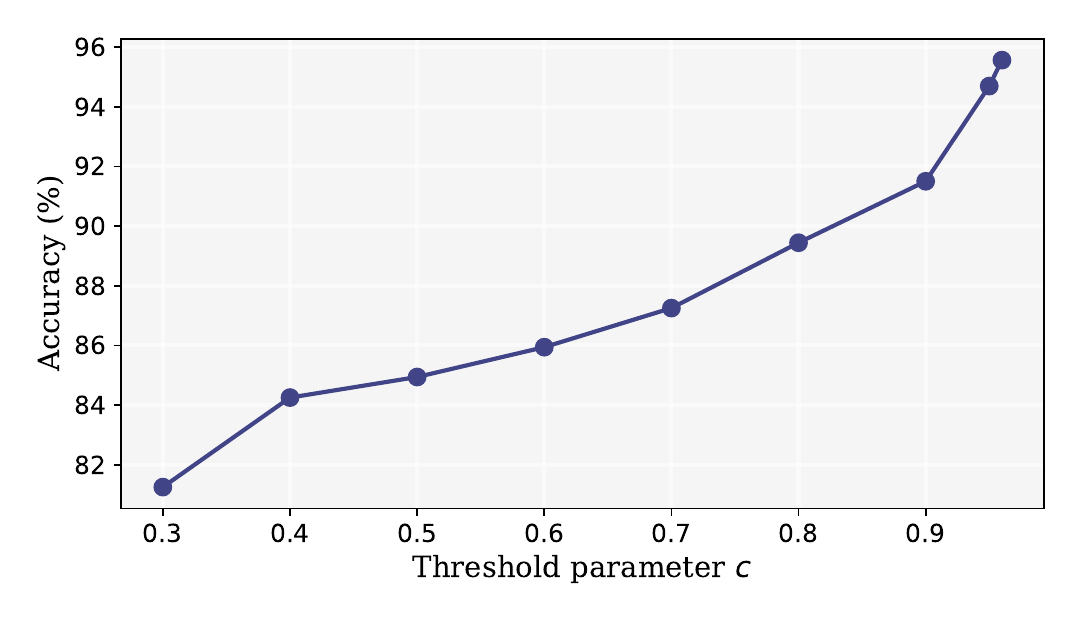}
    \caption{Test accuracy as a function of the threshold parameter \(c\), which defines the boundary between the two bins used to classify Markovian channels based on the memory parameter \(\mu\).}
    \label{fig:results-binning}
\end{figure}
We employ the same RNN architecture described in \cref{app:architectures}. The dataset consists of 8000 samples, each corresponding to a time series of the first component of the covariance matrices of length 10 as input, and the associated class (bin) label as output. The data is split into training and test sets with an 80:20 ratio. To prevent bias, all samples are generated with the same initial mean and variance ($\alpha=\beta=2$).

As shown in \cref{fig:results-binning}, test accuracy increases with the value of \(c\), which is expected since higher values of \(c\) approach the regime of compound channels ($\mu=1$). This behavior suggests that, if \(c\) is interpreted as the threshold separating \textit{low memory} from \textit{high memory} regimes, Markovian channels do not exhibit a memory behavior that scales linearly with the memory parameter \(\mu\). Rather, their effective memory appears to be strongly unbalanced toward \(\mu = 1\). As a result, even relatively high values of \(\mu\) may correspond to channels with weak memory, making them more likely to be misclassified as memoryless.

\paragraph*{Model Complexity.---} In this section, we investigate the impact of model complexity on classification performance. To this end, we train and evaluate five different neural network architectures distinguished primarily by their number of trainable parameters: approximately 1000, 10000, 30000, 100000, and 300000. Our goal is to assess how increasing the model size affects the classification accuracy when distinguishing among quantum channels with different memory properties. 

Table~\ref{tab:classification} reports the classification accuracies achieved by these models on dataset $D_2$ with sequence length 10. The results show that the model with roughly 30000 parameters achieves the best performance, outperforming both smaller and larger architectures. Notably, increasing the number of parameters beyond this point does not lead to further improvement, indicating that excessive model complexity does not necessarily translate into better accuracy. This finding suggests that a moderate model size provides a suitable balance between expressiveness and generalization for the classification task at hand.
\begin{table}[htbp]
\centering
\caption{Classification accuracy on dataset $D_2$ (sequence length 10), as a function of model complexity measured by the number of trainable parameters.}
\label{tab:classification}
\scalebox{1.2}{
\begin{tabular}{cc}
\hline
\textbf{N. of Parameters ($\cdot 10^3$)} & \textbf{Accuracy (\%)} \\
\hline
1    & 59.17 \\
10   & 75.63 \\
30   & \textbf{76.81} \\
100  & 73.89 \\
300  & 73.57 \\
\hline
\end{tabular}}
\end{table}

\subsection{Regression} In the regression task, the objective is to predict the sequence $\{\eta_k\}_{k=1}^5$ corresponding to the input covariance matrices $\{\boldsymbol{\sigma}^{(k)}_r\}_{k=1}^5$. The dataset consists of the same input features as the classification task: a time series but with 5 elements, where each element represents the first diagonal component of the covariance matrix associated with the $k$-th use of the channel. The outputs for the regression are precisely the transmissivity values $\eta_k$ corresponding to the same 5 consecutive channel uses. Therefore, each input sequence of covariance matrix elements corresponds directly to the exact sequence of transmissivity parameters that generated it. The dataset contains 20000 samples, split into training and test subsets with an 80:20 ratio.

The regression model employs a neural network architecture consisting of three layers: a first layer with 64 neurons and ReLU activation, a second layer with 32 neurons and ReLU activation, and a final layer with a number of neurons equal to the sequence length. The model was trained using a learning rate $lr=0.001$ for 200 epochs and a batch size of 1000. 

We opted for a feedforward neural network due to its simplicity and ability to yield good results, making it unnecessary to introduce more complex architectures for this task. For the same reason, we strictly considered the $D_2$ option for the dataset generation, since it is more complicated but the results are good. Similarly, we selected the $D_2$ dataset generation approach, despite its increased complexity, because it still yields excellent results.
\begin{figure}[]
    \centering
    \includegraphics[trim={0.65cm 0.65cm 0.7cm 0.62cm},clip, width=0.47\textwidth]{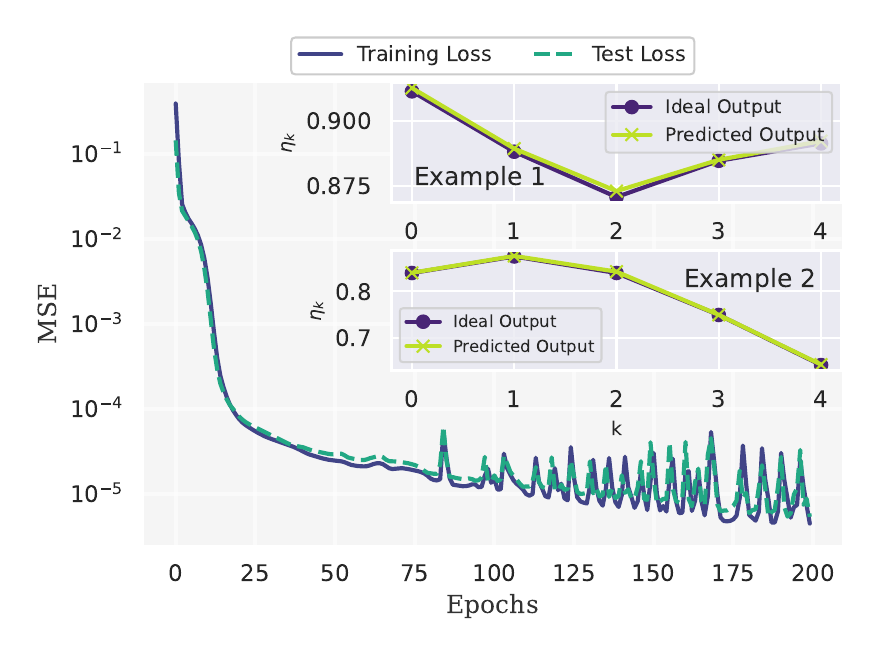}
    \caption{Regression results for predicting the sequence $\{\eta_k\}_{k=1}^5$. The main plot shows the loss function (mean squared error) over training epochs for both the training and test sets. The insets display two representative test samples, comparing the predicted $\eta_k$ values with the corresponding ground truth values.}
    \label{fig:regression}
\end{figure}
The results of the regression task are presented in \cref{fig:regression}, where the mean squared error (MSE) is found to be on the order of $10^{-6}$.

The plot shows the convergence of the loss function for both the training and test sets, confirming the absence of overfitting. Additionally, in the insets, we include two representative test sample predictions, which showcase the model ability to accurately predict the sequence of $\eta_k$ values. These results highlight the network effectiveness in reconstructing the sequence $\{\eta_k\}_{k=1}^5$ that defines the covariance matrix, thereby providing a precise characterization of the underlying quantum channel.
\paragraph*{Model Complexity.---}To investigate the role of model complexity, we trained feedforward neural networks with different numbers of trainable parameters by varying the width of the hidden layers, in order to solve the same regression task addressed before. The corresponding results are reported in Table~\ref{tab:regression}. Notably, a simple architecture with approximately 2000 parameters is sufficient to achieve optimal reconstruction accuracy. Neither reducing the number of parameters to around 1000 nor increasing them to 10000 significantly affects performance, suggesting that the regression task can be effectively solved using compact models with minimal complexity.

\begin{table}[htbp]
\centering
\caption{Mean squared error (MSE) for the reconstruction of transmissivity parameter $\eta_k$ using feedforward neural networks with varying numbers of trainable parameters.}
\label{tab:regression}
\scalebox{1.2}{
\begin{tabular}{cc}
\hline
\textbf{N. of Parameters ($\cdot 10^3$)} & \textbf{MSE} \\
\hline
1    & $10^{-5}$ \\
2    & $10^{-5}$ \\
10   & $10^{-5}$ \\
\hline
\end{tabular}}
\end{table}
\subsection{Forecasting}
Here, we conduct two distinct tests to assess the model ability to predict future values of the lossy parameter \(\eta_{k+1}\) using preceding sequences of the first component of the covariance matrices \(\{\boldsymbol{\sigma}^{(k)}_r\}_k\).
\begin{figure}[]
    \centering
        \includegraphics[trim={0.6cm 0.2cm 0.4 0},clip,width=1\linewidth]{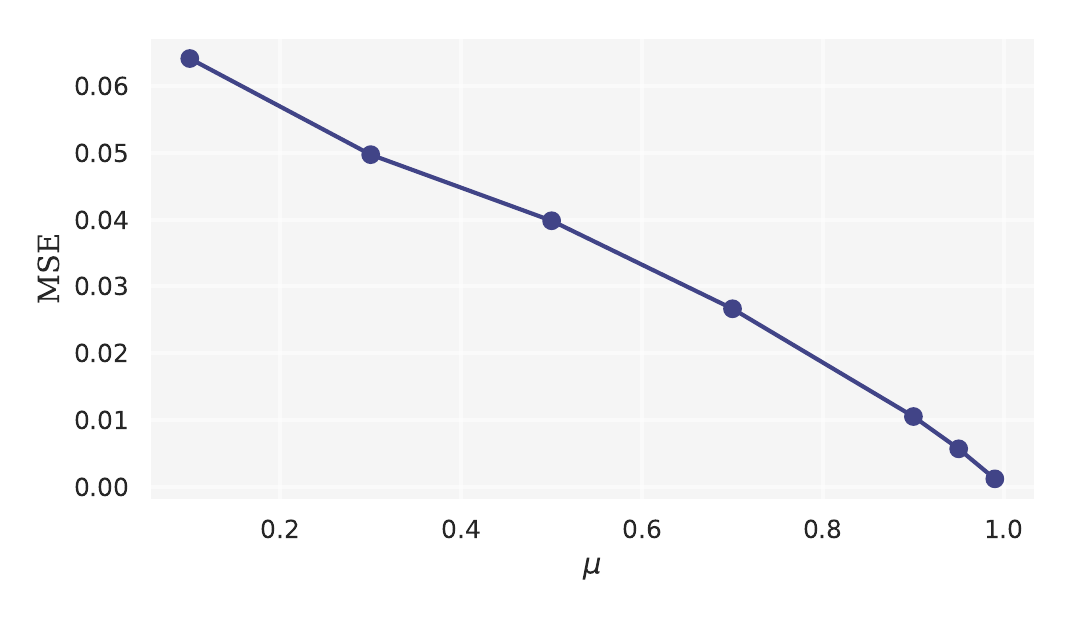}
        \label{fig:subfigfor1}
    \caption{Forecasting error as a function of the memory parameter \(\mu\) for the Markovian process.}
    \label{fig:forecastingmarkovian}
\end{figure}

\begin{figure*}[t]
    \centering
    \subfloat[\label{subfig:a1}]{{\includegraphics[trim={0.6cm 0.6cm 0.2cm 0.6cm},clip, width=0.45 \textwidth]{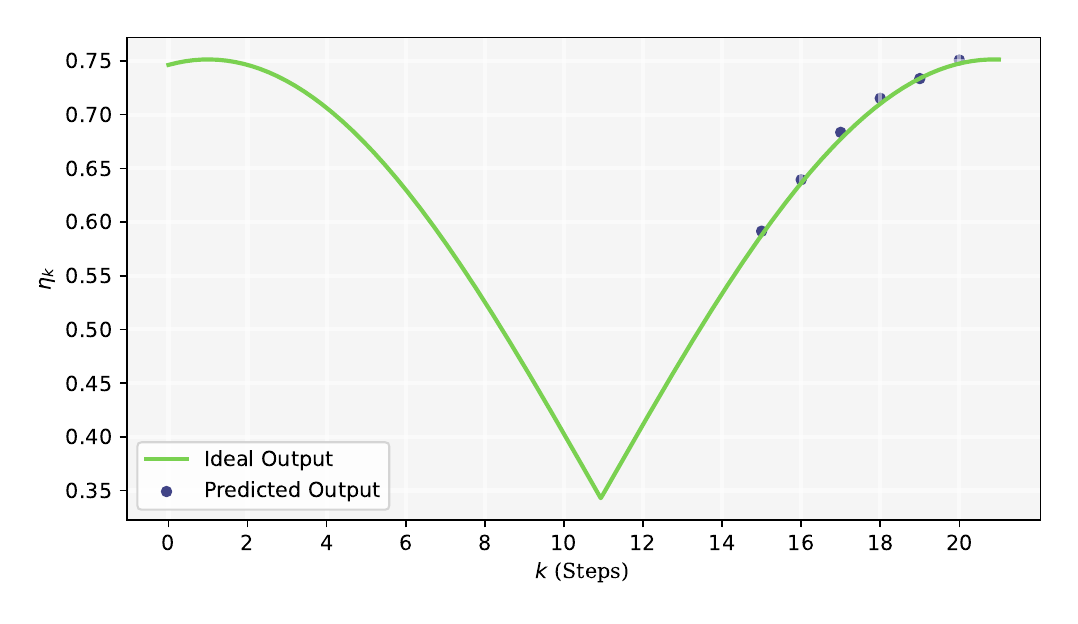} }}%
    \subfloat[\label{subfig:a2}]{{\includegraphics[trim={0.6cm 0.6cm 0.2cm 0.6cm},clip, width=0.45 \textwidth]{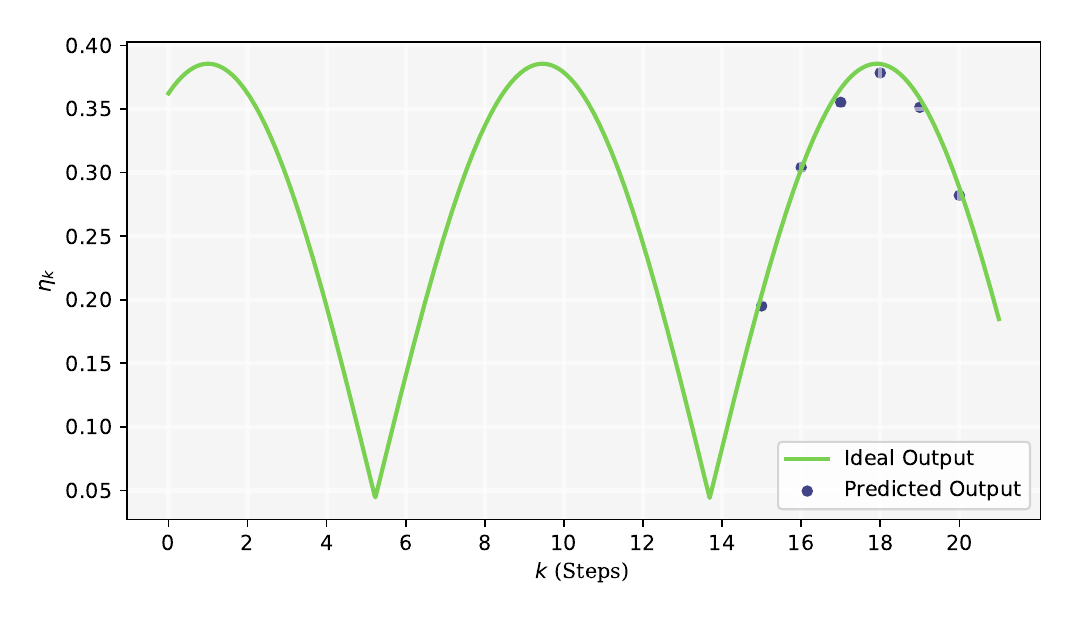} }}\\
    \subfloat[\label{subfig:a3}]{{\includegraphics[trim={0.6cm 0.6cm 0.2cm 0.6cm},clip, width=0.45 \textwidth]{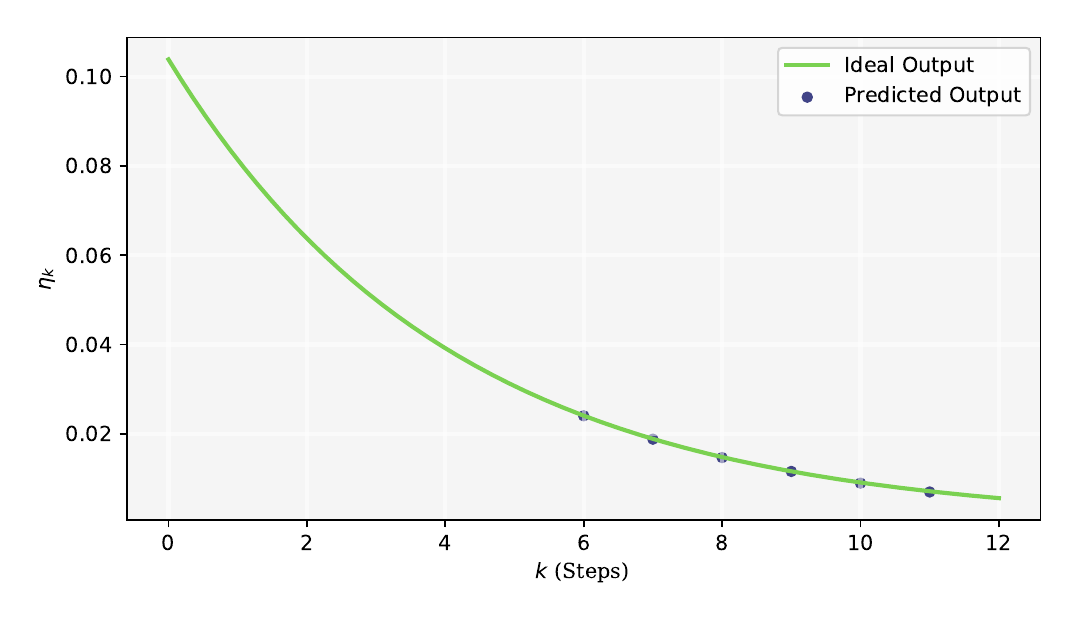} }}%
    \subfloat[\label{subfig:a4}]{{\includegraphics[trim={0.6cm 0.6cm 0.2cm 0.6cm},clip, width=0.45 \textwidth]{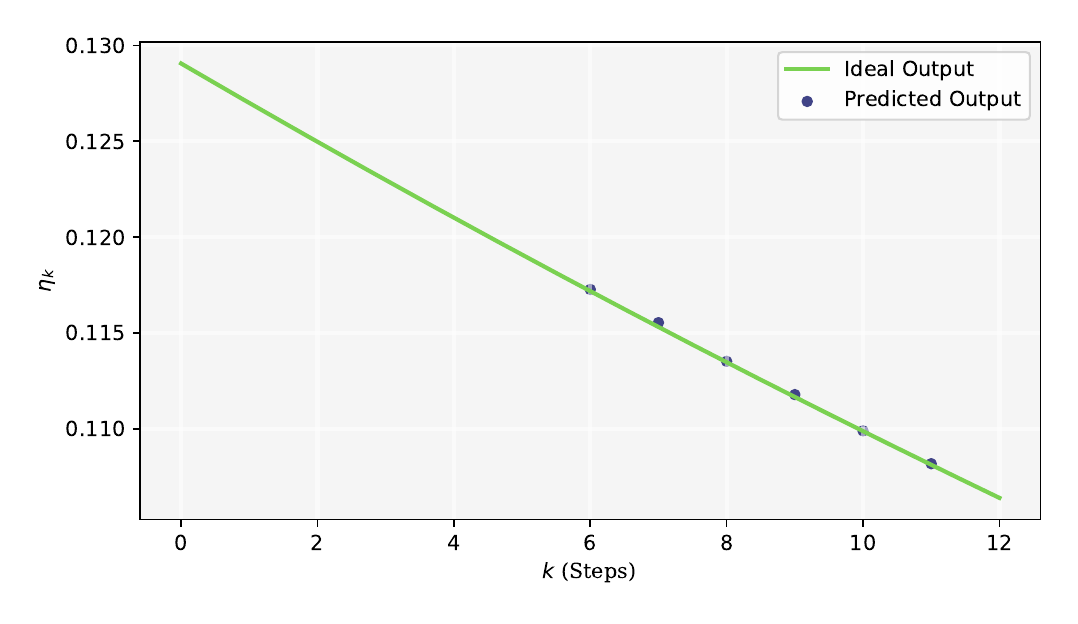} }}%
    \caption{Forecasting 6 future values of \(\eta_k\) from the time series of a component of the covariance matrix for different functional behaviors. (a) and (b) show cosine-like oscillations, while (c) illustrates an exponentially decaying pattern. All cases achieve a mean squared error (MSE) on the order of \(10^{-5}\), indicating high prediction accuracy.}
    \label{fig:forecasting}
\end{figure*}

\paragraph*{Dependence from memory.---} The first test focuses on the Markovian process and investigates how the forecasting performance of the model depends on the memory parameter \(\mu\), which governs the degree of memory in the system. As shown in \cref{fig:forecastingmarkovian}, the mean squared error (MSE) decreases as \(\mu\) approaches 1. This trend indicates that the network more effectively predicts future \(\eta_k\) values when the Markovian process exhibits less memory decay. Notably, even when \(\mu\) takes smaller values, and the samples are drawn randomly from a probability distribution, the forecasting error remains relatively low, demonstrating the model robustness.  

Here, the input consists of the first six steps of the covariance matrix $\{\boldsymbol{\sigma}^{(k)}_r\}_{k=1}^6$, while the target output corresponds to the three subsequent $\{\eta_k\}_{k=7}^9$. The dataset includes 1000 samples, and the data is split into training and test sets in an 80:20 ratio. The neural network architecture is composed of two fully connected layers: the first layer contains 32 neurons, and the second layer contains 16 neurons, both with ReLU activation functions. The final output layer uses a linear activation function to predict the three subsequent \(\eta_k\) values. The model was trained for 500 epochs with a batch size of 100.
\paragraph*{Deterministic channels.---} The second test focuses on evaluating the network forecasting capabilities for deterministic channels. In this case, the input consists of \(K\) steps of the first component of the covariance matrix \(\{\boldsymbol{\sigma}^{(k)}_C\}_{k=1}^K\). The output corresponds explicitly to the next six transmissivity parameter values \(\{\eta_k\}_{k=K+1}^{K+6}\),  which the network aims to predict based on the given input sequence. 

We evaluate the model on two types of processes: cosine-like functions with \(K=15\), and exponential-like functions with \(K=6\). A larger input length is used for the cosine-like case due to its higher complexity, which requires more information to achieve accurate predictions.

The results are shown in \cref{fig:forecasting}, where the prediction for four particular test samples is shown. The model achieves a mean squared error (MSE) of approximately \(10^{-5}\) for both cases. For both tests, the dataset comprises 200000 training samples and 10000 test samples. The neural network architecture features three fully connected layers with 256, 64, and 32 neurons, respectively, each using ReLU activation functions. The model was trained for 500 epochs with a batch size of 1000.

\section{Conclusion}
\label{sec:conclusion}
In this work, we investigated the capabilities of neural networks to analyze, classify, and forecast time-varying quantum channels characterized by their covariance matrices.

For the classification task, we demonstrated that the network effectively distinguishes between non-Markovian, Markovian, memoryless, compound, and deterministic channels, achieving strong performance overall. However, differentiating Markovian from memoryless channels remains challenging, likely because the input features do not always capture sufficiently pronounced correlations. Through systematic experiments varying model complexity, we found that classification performance benefits from moderately complex architectures—around 30,000 trainable parameters—while further increasing the parameter count does not necessarily improve accuracy. Notably, classification requires both a richer architecture, including recurrent layers and dropout techniques, and a larger model size compared to regression and forecasting tasks.

We also recognize limitations related to the current nonlinear modeling of the memory parameter $\mu$ in Markovian channels, which likely affects the network’s ability to clearly distinguish such channels from memoryless ones. Improving this representation, perhaps by introducing a more linear characterization of memory effects, constitutes a promising direction for future work.

For the regression task, which focuses on reconstructing the sequence of transmissivity parameters $\eta_k$, the neural network demonstrated accurate sequence recovery and effective channel characterization. Unlike classification, regression proved to be a simpler problem, where lightweight feedforward neural networks with as few as a couple thousand parameters performed on par with more complex models. Varying model sizes between 1,000 and 10,000 parameters did not yield significant performance differences, indicating that regression task demands considerably less model complexity.

Regarding the forecasting task, the network successfully predicted future values of $\eta_k$ based on past observations, adapting well to different underlying dynamics including Markovian and deterministic processes. Forecasting accuracy notably improved as the memory parameter $\mu$ approached 1, consistent with stronger temporal correlations. Similar to regression, forecasting tasks are effectively addressed by relatively simple architectures.

As a future direction, the methodology could be extended to include a wider variety of quantum channels, such as those with additional noise sources, or even with non-Gaussian noise. This would further evaluate the robustness and scalability of the neural network models in capturing the dynamics of open quantum systems.

\section{Data availability statement}
The data and code that support the findings of this study are publicly available in the GitHub repository~\cite{morgillo2025github}.

\section{Acknowledgments}
A.R.M. and S.M. acknowledge financial support from the PNRR MUR Project PE0000023-NQSTI. M.F.S. and C.M. acknowledge financial support from the PRIN MUR Project 2022SW3RPY.

\appendix

\section{Architectures for classification}
\label{app:architectures}
In this section, we provide an overview of the architectures used for the classification task, along with the associated hyperparameters.
\paragraph*{Random Forest.---} A Random Forest is an ensemble learning method that constructs multiple decision trees during training, and outputs the class that is the mode of the classes (classification) or mean prediction (regression) of the individual trees. We employed a Random Forest classifier with 100 estimators (trees) to ensure robust performance while maintaining computational efficiency.
\paragraph*{Feedforward Neural Network.---} We employed a feedforward neural network architecture consisting of an input layer, followed by a flattening layer, and two hidden layers with 128 neurons each, both using ReLU activation functions. To mitigate overfitting, Dropout layers with a rate of 0.2 were inserted after these layers. Subsequently, two additional hidden layers with 64 neurons and ReLU activation were added, followed by another Dropout layer with the same rate. A further hidden layer with 16 neurons and ReLU activation was then included. The final output layer comprises 5 neurons with a softmax activation function for multi-class classification. The model was trained for 400 epochs using a learning rate of 0.001 and a batch size of 1000.
\paragraph*{Recurrent Neural Network.---} The classification model is based on a Long Short-Term Memory (LSTM) network. It consists of an initial LSTM layer with 64 units and the hyperbolic tangent as the activation function, returning sequences to the next layer. A second LSTM layer with 32 units follows, without returning sequences. Fully connected layers with 64 and 16 neurons, both using ReLU activation, are included, along with Dropout layers (rate 0.2) to prevent overfitting. The final output layer consists of 5 neurons with softmax activation. The model was trained for 800 epochs with a batch size of 1000.
\paragraph*{1D-CNN.---} We employed a one-dimensional Convolutional Neural Network (1D-CNN) that consists of an initial convolutional layer with 128 filters, a kernel size of 2, and ReLU activation, followed by max pooling with a pool size of 2. A second convolutional layer with 128 filters and the same kernel size is applied, followed by another max pooling operation. The feature maps are then flattened and passed through a fully connected layer with 64 neurons using ReLU activation. The final output layer consists of 5 neurons with softmax activation. The model was trained for 400 epochs and a batch size of 1000.

\bibliography{bibliography}
\end{document}